\documentclass[preprint,superscriptaddress,amsmath,amssymb,aps,prl,showkeys]{revtex4-2}
\usepackage{graphicx}
\usepackage{dcolumn}
\usepackage{bm}
\usepackage{amsfonts}
\usepackage{hyperref}
\usepackage[mathlines]{lineno}

\begin{document}

\title{The general propagator for S-wave threshold states}

\author{Hongge Xu}%
\author{Ning Yu}
\email{ning.yuchina@gmail.com}
\author{Zuman Zhang}
\author{Guoying Chen}
\affiliation{School of Physics and Mechanical Electrical \& Engineering, Hubei University of Education, Wuhan 430205, China}
\affiliation{Institute of Theoretical Physics, Hubei University of Education, Wuhan 430205, China}

\date{\today}

\begin{abstract}
We demonstrate that the propagator, derived from an Effective Field Theory (EFT) that incorporates Weinberger's compositeness theorem, provides a more general formula for describing S-wave near-threshold states. By fitting the lineshape using this propagator, we can extract the $Z$ factor for these states and elucidate their structures.   
\end{abstract}
    
\maketitle
\flushbottom   
\section{Introduction}\label{sec1}
The Belle Collaboration first observed the $X(3872)$ state~\cite{Belle:2003nnu}. Since then, an increasing number of nonconventional (exotic) states have been discovered~\cite{particle2022review}. Despite the renewed interest in the spectroscopy of these exotic (or ${XYZ}$) states over the past two decades~\cite{Esposito:2021vhu,Chen:2022asf}, their nature remains uncertain.

Effective field theory is widely used in hadron physics, and it is also used for studying the exotic states. For example in Ref~\cite{Fleming:2007rp}, an effective field theory, which is called XEFT, is proposed to study $X(3872)$. In the XEFT, the $X(3872)$ is assumed to be a weakly bound molecule of charm meson pair. Recently, the LHCb reports the observation of decay mode ${X(3872)}\rightarrow\psi(2S)\gamma$, and the ratio of this partial width to that of ${X(3872)}\rightarrow J/\psi\gamma$ is measured to be  $R_{\psi\gamma}=\frac{\Gamma_{X(3872)\rightarrow\psi(2S)\gamma}}{\Gamma_{X(3872)\rightarrow J/\psi\gamma}}=1.67\pm0.21\pm0.12\pm0.04$~\cite{LHCb:2024tpv}. The value $R_{\psi\gamma}$ is sensitive to the nature of $X(3872)$. The measured ratio makes the interpretation of $X(3872)$ as pure $D\bar D^*$ molecule questionable and indicates sizeable compact component in $X(3872)$~\cite{LHCb:2024tpv}. In view of this, it is necessary to consider the compact component in effective field theory. In Ref~\cite{Chen:2013upa}, an effective field theory which incorporates Weinberg's compositeness theorem and considers the $X(3872)$ as a compact state which couple strongly to $D\bar D^*$ is proposed. To distinguish such an effective field theory from the XEFT, we will call the effective field theory proposed in Ref~\cite{Chen:2013upa} as CEFT throughout this paper. The CEFT is further used to study the structure of the $Z_b$ states~\cite{Huo:2015uka} and very recently to study the structure of $X(3872)$~\cite{Duan:2024zuo,Xu:2023lll}. The aim of the paper is to discuss some property of CEFT, in particularly, the propagator function in CEFT.
    
Typically, three amplitudes are used to describe these near-threshold states: Breit-Wigner, Flatt$\acute{e}$ and low-energy scattering amplitude. Let's consider a two-body channel, denoted as $DD$, with a threshold $M_{th}$ and a near-threshold state ${X}$ with mass $M$ and width $\Gamma$. The Breit-Wigner propagator is defined as $\dfrac{i}{ P^2-M^2+iM\Gamma}$, where $P^\mu=({E_0},\overrightarrow{P})$. In the center-of-mass frame, $\overrightarrow{P}=(0,0,0)$, ${E_0}$ and ${M}$ can be parameterized as ${E_0}={M_{th}}+E$ and ${M}={M_{th}}-B$, respectively, where $E$ is the energy relative to the threshold, $B$ is the binding energy (we call it binding energy in the sense that it is defined relative to the threshold). We choose the convention that $B>0$ if the state locates below threshold. Near the threshold, terms suppressed by $\dfrac{E}{2{M_{th}}}$ and $\dfrac{B}{2{M_{th}}}$ can be neglected. The amplitude can then be written as $f(E)=\dfrac{1}{D_{BW}(E)}$, where
\begin{equation}
D_{BW}(E)=E+B+i\Gamma/2. \label{sec1:1}
\end{equation}

The Flatt$\acute{e}$ formula is a general model used to parameterize the resonant structure near a hadron-hadron threshold~\cite{Flatte:1976xu}. The Flatt$\acute{e}$ amplitude can be written as $f(E)=\dfrac{1}{D_{FL}(E)}$, with $D_{FL}(E)$ defined as
\begin{equation}
D_{FL}(E)=E-E_f-\frac{1}{2}g_1\sqrt{-2\mu{E}-i\epsilon}+i\frac{1}{2}\Gamma_f.  \label{sec1:2}
\end{equation}
Here, $E_f$ is the Flatt$\acute{e}$ energy parameter related to the mass, $\mu$ is the reduced mass of the two-body state $DD$, $g_1$ characterizes the coupling between $X$ and $DD$, and $\Gamma_f$ accounts for decay modes other than $DD$. As we will illustrate later, the Flatt$\acute{e}$ amplitude assumes the existence of a compact object.

Finally, the low-energy amplitude is written as $f(E)=\dfrac{1}{D_{LE}(E)}$, with $D_{LE}(E)$ defined as~\cite{Braaten:2007dw}
\begin{equation}
D_{LE}(E)=-{1}/{a}+\sqrt{-2\mu{E}-i\epsilon},   \label{sec1:3}
\end{equation}
where $a$ is the scattering length. We call it low-energy amplitude, one should note to distinguish it from the low energy expansion amplitude $f=\frac{1}{-\frac{1}{a}+\frac{1}{2}rp^2-ip}$, where $r$ is the effective range.
    
In this paper, we investigate the relationship between the propagator derived from CEFT and the three amplitudes commonly used. We find that the CEFT-derived propagator serves as a more general form, making it particularly useful for characterizing the lineshape of S-wave near-threshold states.

\section{Revisit Weinberg's compositeness Theorem} \label{sec2}
To proceed, we will first revisit Weinberg's method~\cite{Weinberg:1962hj,weinberg1965evidence}. The total Hamiltonian $H$ of interest can be divided into a free part $H_0$ and an interaction part $V$ 
\begin{equation}
    H=H_0+V.
 \end{equation}
Specifically, $H_0$ can be Hamiltonian in the quark models, where only the interactions between the quarks are considered. The eigenstates of the free part $H_0$ include continuum states $|\alpha\rangle$ (for example, the two-body state $DD$) and the possible discrete bare elementary $|\mathcal{B}\rangle$ near the $DD$ threshold. $|\mathcal{B}\rangle$ is a compact state, possibly a tetraquark state or a quarkonium. $V$ describes the interaction between $|\mathcal{B}\rangle$ and $DD$. We then have
\begin{eqnarray}
H_0|\alpha\rangle&=&E(\alpha)|\alpha\rangle, \quad
\langle\beta|\alpha\rangle=\delta(\beta-\alpha), \quad
H_0|\mathcal{B}\rangle=-B_0|\mathcal{B}\rangle,  \nonumber\\
\langle\alpha|\mathcal{B}\rangle&=&0, \qquad\qquad
\langle\mathcal{B}|\mathcal{B}\rangle=1, \qquad\qquad
\langle\alpha|V|\mathcal{B}\rangle=g_0,  \label{sec2:2}
\end{eqnarray}
where the energies are defined relative to the $DD$ threshold. We adopt the convention that $B_0>0$ if the mass of $|\mathcal{B}\rangle$ is below the $DD$ threshold. We parameterize the matrix element $\langle\alpha|V|\mathcal{B}\rangle$ as a constant $g_0$, since the coupling between $|\mathcal{B}\rangle$ and $|\alpha\rangle$ is an S-wave coupling. Higher-order correction in effective field theory can be considered by replacing $g_0$ with $g_0+g_1p^2$, where $g_1$ is a constant and $p$ is momentum of $D$ in the center-of-mass-frame of $DD$ system. Near the threshold, $p$ is a small scale, allowing us to neglect the term $g_1p^2$ in the leading order result. A physical bound state $|X\rangle$ is a normalized eigenstate of the total Hamiltonian $H$, with
\begin{equation}
H|X\rangle=-B|X\rangle, \quad \langle X|X\rangle=1.
\end{equation}
Without losing generality, the wave function of $|X\rangle$ can be written as 
\begin{equation}
|X\rangle= \sqrt{Z}|\mathcal{B}\rangle+\int d\alpha C_{\alpha}|\alpha\rangle, \label{sec2:04}
\end{equation}
where $Z$ is the probability of finding $|\mathcal{B}\rangle$ in the bound state $|X\rangle$. We refer to $|X\rangle$ as a bound state in the sense that it includes continuum states in its wave function, and its mass is below the threshold, or equally, $B>0$. If $Z=0$, then $|X\rangle$ contains only the continuum states in its wave function, and we will refer to such a state as pure molecular state to distinguish it from a bound state $|X\rangle$ with $Z\neq0$. The coefficient $C_{\alpha}$ can be explored as 
\begin{equation}
C_{\alpha}=\langle\alpha|X\rangle=\langle\alpha|\frac{V}{H-H_0}|X\rangle=-\frac{\langle\alpha|V|X\rangle}{E(\alpha)+B}.  \label{sec2:5}
\end{equation}
Using the wave function of $|X\rangle$, the matrix element $\langle\alpha|V|X\rangle$ in the above can be expressed as
\begin{equation}
\langle\alpha|V|X\rangle=\sqrt{Z}\langle\alpha|V|\mathcal{B}\rangle+\int d\beta C_{\beta}\langle\alpha|V|\beta\rangle,
\end{equation}
note that $\langle\alpha|V|\mathcal{B}\rangle=g_0$, as defined in Eq.\eqref{sec2:2}. If we parameterize the matrix element $\langle\alpha|V|X\rangle$ as what Weinberg had done, $\langle\alpha|V|X\rangle=g$, where $g$ is a constant. By setting $\langle\alpha|V|\beta\rangle=0$, we immediately obtain the relation $g=\sqrt{Z}g_0$, or equally
\begin{equation}
g_0^2={g^2}/{Z}. \label{sec2:07}
\end{equation}
Actually, the above relation is one of the matching relations given in Ref~\cite{Chen:2013upa}. There, the relation is obtained through the matching between a non-relativistic effective field theory and Weinberg's compositeness theorem. This relation can be reproduced by setting $\langle\alpha|V|\beta\rangle=0$, which simply reflects the fact that the direct four-body $DD-DD$ contact interaction terms and also the $D-D$ interactions mediated by the t-channel meson exchange are not explicitly included in the effective field theory of Ref~\cite{Chen:2013upa}. This is because their contributions are of the order $\mathcal{O}(p^0)$ ($p$ is momentum of $D$ )or higher, which is smaller than the leading order $D-D$ scattering amplitude (which is $\mathcal{O}(p^{-1})$) in CEFT. The contribution of the matrix element $\langle\beta|V|\alpha\rangle$ can be taken into account in the next leading order results of CEFT by including the four-body $DD-DD$ contact interaction terms and also the t-channel meson exchange interactions. 
With $\langle\alpha|V|X\rangle=g$, we can write Eq.\eqref{sec2:5} as
\begin{equation}
C_{\alpha}=-\frac{g}{E(\alpha)+B}. \label{sec2:3}
\end{equation}
Using the normalization condition of $|X\rangle$, i.e., $\int d\alpha{|C_{\alpha}|}^2=1-Z$, with
\begin{equation}
d\alpha=\frac{d^3p}{{(2\pi)}^3}=\frac{4\pi p^2dp}{{(2\pi)}^3}=\frac{\mu^{3/2}}{\sqrt{2}\pi^2}E^{1/2}dE, E\equiv{p^2}/{2\mu}, \label{sec2:4}
\end{equation}
we then obtain 
\begin{equation}
g^2=\frac{2\pi\sqrt{2\mu B}}{\mu^2}(1-Z). \label{sec2:1}
\end{equation}
With Eq.\eqref{sec2:1}, Weinberg derive the well-known relations between $Z$, $B$ and effective range expansion parameters, i.e., the scattering length $a$ and effective range $r$, as
\begin{equation}
a=[2(1-Z)/(2-Z)]/\sqrt{2\mu{B}}+O({m_\pi}^{-1}) \label{sec:2},
\end{equation}
\begin{equation}
r=-[Z/(1-Z)]/\sqrt{2\mu{B}}+O({m_\pi}^{-1})\label{sec:3}.
\end{equation}
It is worth mentioning that in the case of $X(3872)$, Ref\cite{Esposito:2023mxw} finds that the corrections to the low-energy expansion parameters from pion are extremely small (see also Ref\cite{Braaten:2020nmc}). Thus, it is a very good approximation to neglect the correction terms $\mathcal{O}({m_\pi}^{-1})$ in Eq.\eqref{sec:2} and Eq.\eqref{sec:3}. For the $X(3872)$, the charged $DD$ channel may also need to be taken into account besides the neutral channel. We will denote the charged continuum state as $|\alpha_c\rangle$, while $|\alpha\rangle$ denotes the neutral continuum $DD$ states. We then have
\begin{equation}
\begin{aligned}
&H_0|\alpha_c\rangle=(E(\alpha_c)+\delta)|\alpha_c\rangle, 
&\langle\alpha|\alpha_c\rangle=0, \qquad&\langle\mathcal{B}|\alpha_c\rangle=0,  \\ 
&\langle\beta_c|\alpha_c\rangle=\delta(\beta_c-\alpha_c),\qquad &\langle\alpha_c|V|\mathcal{B}\rangle=g_{0c},  \label{sec2:13}
\end{aligned}
\end{equation}
where $\delta$ is the mass splitting between the charged channel and neutral channel. $\delta$ is included  because $E(\alpha_c)$ is defined relative to the threshold of the charged channel, while the eigenvalue of $H_0$ is defined relative to the threshold of the neutral channel. We use a new constant $g_{0c}$, which may differ from $g_0$ defined in Eq.\eqref{sec2:2}, to parameterize the matrix element $\langle\alpha_c|V|\mathcal{B}\rangle$, as the isospin violation is sizeable for the $X(3872)$. With the charged channel be taken into account, the wave function of the $X(3872)$ can be written as 
\begin{equation}
|X(3872)\rangle=\sqrt{Z}|\mathcal{B}\rangle+\int d\alpha C_{\alpha}|\alpha\rangle+\int d\alpha_c C_{\alpha_c}|\alpha_c\rangle. \label{sec2:14}
\end{equation}
Along the same line to treat $C_{\alpha}$, we can have
\begin{eqnarray}
C_{\alpha_c}&=&\langle\alpha_c|X(3872)\rangle \nonumber\\
&=&\langle\alpha_c|\frac{V}{H-H_0}|X(3872)\rangle \nonumber\\
&=&-\frac{\langle\alpha_c|V|X(3872)\rangle}{(E(\alpha_c)+\delta)+B} \nonumber\\
&=&-\frac{g_c}{E(\alpha_c)+\delta+B}, \nonumber\\ \label{sec2:9}
\end{eqnarray}
where, we have used the wave function of the $X(3872)$ in Eq.\eqref{sec2:14}, $\langle\alpha_c|V|\mathcal{B}\rangle=g_{0c}$ in Eq.\eqref{sec2:13} and $\langle\alpha_c|V|\alpha\rangle=\langle\alpha_c|V|\beta_c\rangle=0$ in the last step. Thus, the matrix element $\langle\alpha_c|V|X(3872)\rangle$ can be expressed as
\begin{equation}
\langle\alpha_c|V|X(3872)\rangle=\sqrt{Z}\langle\alpha_c|V|\mathcal{B}\rangle=\sqrt{Z}g_{0c}\equiv g_c. \label{sec2:10}
\end{equation}
Again the matrix elements $\langle\alpha_c|V|\alpha\rangle$ and $\langle\alpha_c|V|\beta_c\rangle$ are setting to be zero, as their effects will only appear in the next to leading order results. Now the wave function normalization of $X(3872)$ reads
\begin{equation}
Z+\int d\alpha {|C_\alpha|}^2+\int d\alpha_c {|C_{\alpha_c}|}^2=1.
\end{equation}
After doing the phase space integral, we have
\begin{equation}
g^2\frac{\mu^2}{2\pi\sqrt{2\mu B}}+g_c^2\frac{\mu_c^2}{2\pi\sqrt{2\mu_c(B+\delta)}}=1-Z, \label{sec2:18}
\end{equation}
where, $\mu$ is reduce mass of neutral $DD$, $\mu_c$ is the reduced mass of charged $DD$. Eq.\eqref{sec2:18} is the extension of Eq.\eqref{sec2:1} to further include another continuum spectrum. Note that if we further assume $g^2=g_c^2$ we can obtain the expression for $g^2$ with Eq.\eqref{sec2:18}.

We will further discuss $X(3872)$ latter, and now we will turn to study the matrix element $\langle\mathcal{B}|X\rangle$, we have
\begin{equation}
\sqrt{Z}=\langle\mathcal{B}|X\rangle=\langle\mathcal{B}|\frac{V}{H-H_0}|X\rangle=\frac{\langle\mathcal{B}|V|X\rangle}{-B+B_0},
\end{equation}
or equally
\begin{equation}
\langle\mathcal{B}|V|X\rangle=\sqrt{Z}(B_0-B).  \label{sec2:7}
\end{equation}
On the other hand, $\langle\mathcal{B}|V|X\rangle$ can also be estimated by using the wave function of $X$, i.e., Eq.\eqref{sec2:04} as
\begin{equation}
\langle\mathcal{B}|V|X\rangle=\int d \alpha C_{\alpha}\langle\mathcal{B}|V|\alpha\rangle.
\end{equation}
Note that we set $\langle\mathcal{B}|V|\mathcal{B}\rangle=0$, since at leading order $V$ describes the interactions between $|\mathcal{B}\rangle$ and $DD$. The above integral has ultraviolet divergence and can be handled in the momentum space with dimensional regularization. Using the definition $\langle\alpha|V|\mathcal{B}\rangle=g_0$ in Eq.\eqref{sec2:2}, and relations in Eq.\eqref{sec2:07}, \eqref{sec2:3} and  \eqref{sec2:4}, we have
\begin{eqnarray}
\langle\mathcal{B}|V|X\rangle&=&\int d \alpha C_{\alpha}\langle\mathcal{B}|V|\alpha\rangle\nonumber\\
&=&-\frac{g^2}{\sqrt{Z}}\int\frac{d^3p}{(2\pi)^3}\frac{1}{p^2/(2\mu)+B} \nonumber\\
&=&-\frac{g^2}{\sqrt{Z}}\nu^{4-D}\int\frac{d^{D-1}p}{(2\pi)^{D-1}}\frac{1}{p^2/(2\mu)+B} \nonumber\\
&=&-\frac{g^2}{\sqrt{Z}}\nu^{4-D}2\mu\frac{\Gamma(\frac{3-D}{2})}{{(4\pi)}^{\frac{D-1}{2}}\Gamma(1)}{(2\mu B)}^{\frac{D-3}{2}},  \label{sec2:22}
\end{eqnarray}
where, $\nu$ is the additional scale introduced in dimensional regularization. Note that there is no pole in $D=4$. After takeing the $D\rightarrow4$ limit and using Eq.\eqref{sec2:1}, we obtains
\begin{equation}
\langle\mathcal{B}|V|X\rangle=\frac{g^2}{\sqrt{Z}}\frac{\mu}{2\pi}\sqrt{2\mu B}=2\frac{1-Z}{\sqrt{Z}}B. \label{sec2:8}
\end{equation}
Combined Eq.\eqref{sec2:7} and Eq.\eqref{sec2:8}, we have
\begin{equation}
B_0=\frac{2-Z}{Z}B.  \label{sec2:24}
\end{equation}
Which is exactly the other matching relations that was obtained in Ref~\cite{Chen:2013upa}. Here we reproduced it using a different approach as a cross-check. With the new approach, we can straightforwardly extend the above relation to $X(3872)$, where the charged channel may also be included. Using the wave function for the $X(3872)$ in Eq.\eqref{sec2:14}, $C_{\alpha_c}$ in Eq.\eqref{sec2:9}, the matrix element $\langle\alpha_c|V|\mathcal{B}\rangle$ in Eq.\eqref{sec2:10}, we obtain
\begin{equation}
\begin{aligned}
\langle\mathcal{B}|V|X(3872)\rangle=\int d \alpha C_{\alpha}\langle\mathcal{B}|V|\alpha\rangle+\int d \alpha_c C_{\alpha_c}\langle\mathcal{B}|V|\alpha_c\rangle  \\
=\frac{g^2}{\sqrt{Z}}\frac{\mu}{2\pi}\sqrt{2\mu B}+\frac{g_c^2}{\sqrt{Z}}\frac{\mu_c}{2\pi}\sqrt{2\mu_c(B+\delta)}. \label{sec2:11}
\end{aligned}
\end{equation}
The integrals in the above are treated in the same way as in Eq.\eqref{sec2:22}. Combining Eq.\eqref{sec2:11} and Eq.\eqref{sec2:7}(note that Eq.\eqref{sec2:7} now becomes $\langle\mathcal{B}|V|X(3872)\rangle=\sqrt{Z}(B_0-B)$), we can then generalize Eq.\eqref{sec2:24} to
\begin{equation}
B_0=\frac{1}{Z}[g^2\frac{\mu}{2\pi}\sqrt{2\mu B}+g_c^2\frac{\mu_c}{2\pi}\sqrt{2\mu_c(B+\delta)}+ZB].  \label{sec2:26}
\end{equation}
We will utilize this relation later to obtain the propagator for the $X(3872)$. Before ending this section, we would like to discuss the interpretation of $Z$ in Weinberg's compositeness theorem. As previously illustrated, $Z$ represents the probability of finding $|\mathcal{B}\rangle$ in a physical bound state $|X\rangle$. However, $|\mathcal{B}\rangle$ is not a physical state but rather an the eigenstate of $H_0$, with the interaction $V$ turned off. When the interaction $V$ turns on, $|\mathcal{B}\rangle$ will couple with the continuum state $DD$, resulting $0<Z<1$. Consequently, a large $Z$, for example close to one, indicates that the compact state couples weakly with the continuum state. Conversely, a small but non-vanishing $Z$ indicates that the compact state couples strongly with the continuum state. On the other hand, $Z=0$ implies that the physical state $|X\rangle$ is dynamically generated by the $DD$ interaction without a compact state. It is noting that this point has also been addressed in Ref\cite{Esposito:2021vhu}.


\section{The general propagator for the S-wave near threshold states}

Weinberg's compositeness theorem has been incorporated into a non-relativistic effective field theory (CEFT) in Ref~\cite{Chen:2013upa}. There, the propagator for the S-wave near threshold state is written as 
\begin{equation}
G_X(E)=\frac{iZ}{D_{EFT}(E)},  \label{sec3:1}
\end{equation}
\begin{equation}
D_{EFT}(E)=E+B+\widetilde{\Sigma^\prime}(E)+i\Gamma/2,
\end{equation}
\begin{equation}
\widetilde{\Sigma^\prime}=-g^2[\frac{\mu}{2\pi}\sqrt{-2\mu{E}-i\epsilon}+\frac{\mu\sqrt{2\mu{B}}}{4\pi{B}}(E-B)].
\end{equation}
The term $i\Gamma/2$ was directly inserted into the denominator of the above propagator, thereby rendering the physical interpretation of $\Gamma$ ambiguous~\cite{Chen:2013upa}. Now, we will re-derive the propagator and find that the $\Gamma$ can be naturally incorporated in Field theory. This methodology will elucidate that the $\Gamma$ has a distinct physical interpretation, namely, it is the renormalized width resulting from the non-$DD$ decays of $X$.
%

Consider a bare state $|\mathcal{B}\rangle$ with bare mass $-B_0$, width $\Gamma_0$, and coupling $g_0$ to the two-particle state $DD$. If $|\mathcal{B}\rangle$ is near the two-particle threshold, the propagator of this bare state can be written as $\dfrac{i}{E+B_0+i\Gamma_0/2}$ as shown in Eq.\eqref{sec1:1}. The full propagator can be obtained by summing the Feynman diagrams in Fig. 1. Near threshold, the momenta of the particles are non-relativistic. The loop integral in Fig.1 can be calculated straightforwardly with the minimal subtraction (MS) scheme, the result can be written as ~\cite{Kaplan:1998tg}
\begin{align}
\mathcal{I}^{MS}&\equiv\nu^{4-D}\int\frac{d^D\ell}{(2\pi)^D}\frac{i}{[\ell^0-\vec{\ell}^2/(2m_1)+i\epsilon]} \nonumber\\
&\quad{}\times\frac{i}{[E-\ell^0-\vec{\ell}^2/(2m_2)+i\epsilon]}\nonumber\\
&=i\nu^{4-D}\int\frac{d^{D-1}\ell}{(2\pi)^{D-1}}\frac{1}{E-\frac{\vec{\ell}^2}{2\mu}+i\epsilon} \nonumber\\
&=-i\nu^{4-D}\frac{2\mu}{(4\pi)^{\frac{D-1}{2}}}{(-2\mu E-i\epsilon)}^{\frac{D-3}{2}}\Gamma(\frac{3-D}{2}) \nonumber\\
&=i\frac{\mu}{2\pi}\sqrt{-2\mu E-i\epsilon}. \label{sec3:4}
\end{align}
\begin{figure}[htb]
\includegraphics[width=0.85\textwidth]{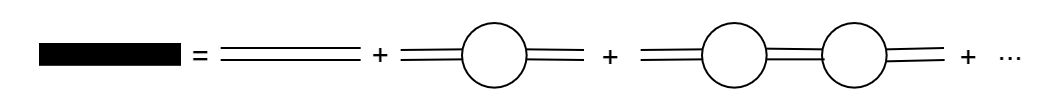}
\caption{Full propagator for the near threshold state. The double line denotes the bare state.}
\label{tu2}
\end{figure}
Where $m_1$ and $m_2$ are the masses of $DD$ states. Note that the above loop integral is linearly divergent and has a pole at $D=3$, but it has no pole at $D=4$. In the minimal subtraction scheme, counter terms are added to subtract the pole at $D=4$. Since the result has no pole at $D=4$, no counter term is needed in $MS$ scheme.
It is easy to consider the widths of $DD$ states by replacing the propagator in Eq.\eqref{sec3:4} with $\frac{i}{\ell^0-\vec{\ell}^2/(2m)+i\Gamma/2}$. The result can be obtained by simply replacing the $\epsilon$ in the final expression of Eq.\eqref{sec3:4} with $\mu(\Gamma_1+\Gamma_2)$, where $\Gamma_1$ and $\Gamma_2$ are the decay widths of $DD$ states (Similar result is also given in Ref~\cite{Braaten:2007dw} when $X$ is a pure molecular state.). It is worth mentioning that due to the absence of log divergences, the additional renormalization scale does not appear in the final result of loop integral. This feature makes it possible to incorporate the relation Eq.\eqref{sec2:1}, which is obtained from quantum mechanics, into the EFT (as in Ref~\cite{Chen:2013upa}). Because if the result of the loop integral depends on the renormalization scale, the coupling $g^2$ will also depend on the renormalization scale through the renormalization group equation. Similar argument can be applied to the field renormalization constant $Z$, thus $Z$ does not depend on the renormalization scale and can have physical interpretation, i.e., the probability.

After summing the contributions from all the diagrams, the full propagator reads
\begin{equation}
i\varDelta=\frac{i}{E+B_0-g_0^2\frac{\mu}{2\pi}\sqrt{-2\mu{E}-i\epsilon}+i\Gamma_0/2}. \label{sec3:5}
\end{equation}

Actually, the above amplitude is just the Flatt$\acute{e}$ amplitude. The separation of the self energy into two terms $i\Gamma_0/2$ and $-g_0^2\frac{\mu}{2\pi}\sqrt{-2\mu{E}-i\epsilon}$ may not make too much sense if the mass of the bare state is well above the threshold, because in this case $-g_0^2\frac{\mu}{2\pi}\sqrt{-2\mu{E}-i\epsilon}$ is pure imaginary as the $i\Gamma_0/2$ term in the concerned energy region. But such separation is necessary if the mass of the bare state is below the threshold, as $-g_0^2\frac{\mu}{2\pi}\sqrt{-2\mu{E}-i\epsilon}$ is real for $E<0$~\cite{Flatte:1976xu}. 

To incorporate Weinberg's compositeness theorem, one should use the matching relations between the bare parameters ($g_0$, $B_0$) and the renormalized parameters ($g$, $B$) as given in Ref~\cite{Chen:2013upa}(and they are also reproduced in previous section)
\begin{equation}
g_0^2=g^2/Z,  \label{sec3:6}
\end{equation}
\begin{equation}
B_0=\frac{2-Z}{Z}B, \label{sec3:7}
\end{equation}
where, $g^2$ is defined in Eq.\eqref{sec2:1}, $B$ is the binding energy. Note that $g_0$, $B_0$ become infinite in the limit $Z\rightarrow0$. The matching relation Eq.\eqref{sec3:6} is obtained in the limit $\Gamma_0\rightarrow0$ in Ref~\cite{Chen:2013upa}. We assume that this relation still holds for a non-vanishing $\Gamma_0$, as the long-distant physics (the coupling between $X$ and $DD$ channel ) should not be sensitive to the short-distant physics (the coupling between $X$ and non-$DD$ channel). Notice that $B$ and $B_0$ have the same sign for $0\le{Z}\le1$, so the bare state is below the threshold if $X$ is below the threshold, and vice versa.

Using Eq.\eqref{sec3:6}, Eq.\eqref{sec3:7} and Eq.\eqref{sec2:1}, the full propagator can be rewritten as
\begin{eqnarray}
i\varDelta &=&\frac{iZ}{ZE+(2-Z)B-g^2\frac{\mu}{2\pi}\sqrt{-2\mu{E}-i\epsilon}+iZ\Gamma_0/2}  \nonumber\\
&=&\frac{iZ}{E+B-g^2\frac{\mu}{2\pi}\sqrt{-2\mu{E}-i\epsilon}-(1-Z)(E-B)+iZ\Gamma_0/2}   \nonumber\\
&=&\frac{iZ}{E+B-g^2\frac{\mu}{2\pi}\sqrt{-2\mu{E}-i\epsilon}-(1-Z)\frac{2\pi\sqrt{2\mu B}}{\mu^2}\cdot\frac{\mu^2}{2\pi\sqrt{2\mu B}}(E-B)+iZ\Gamma_0/2}   \nonumber\\
&=&\frac{iZ}{E+B-g^2\frac{\mu}{2\pi}\sqrt{-2\mu{E}-i\epsilon}-g^2\frac{\mu\sqrt{2\mu B}}{4\pi B}(E-B)+iZ\Gamma_0/2},  \label{sec3:8}
\end{eqnarray}
which is precisely the form of $G_X(E)$ as defined in Eq.\eqref{sec3:1}. By comparing Eq.\eqref{sec3:1} and Eq.\eqref{sec3:8}, we can find that $\Gamma$ satisfies $\Gamma=Z\Gamma_0$. Thus, we have demonstrated that the width $\Gamma$ can be naturally incorporated into CEFT. This width can be interpreted as the renormalized width, which comes from the non-$DD$ decay of the elementary state. In this manner, one can naturally extend Weinberg's compositeness theorem to resonances. Actually, what we have addressed above is the coupled channel effect between the $DD$ decay channel and the non-$DD$ channels. The parameters $B_0$ and $\Gamma_0$ may have physical interpretation, in other words, they may be calculated from some specific tetraquark model. For $0<Z<1$, the coupled channel effect will shift the mass of the state near to the threshold, i.e., $B<B_0$ in Eq.\eqref{sec3:7}.

We now come to discuss the relation between the propagator $G_X(E)$ and the Breit-Wigner, Flatt$\acute{e}$, low-energy amplitude. Firstly, for $Z=1$ ($g^2 = 0$), one can find that $D_{EFT}(E)$ is just $D_{BW}(E)$ defined in the near threshold form of Breit-Wigner, i.e., Eq.\eqref{sec1:1}. This can be easily understood as by using Breit-Wigner for the propagator of the bare state in Fig1, we assume the interactions $V$ have been turned off. If the interactions have been turned on ($g^2\neq0$), the propagator of the compact state should be replaced with the bubble sum shown in Fig1. In another word, the Breit-Wigner amplitude assumes the below threshold compact state does not couple with the continuum state. Note that this conclusion can not be extended to a compact state with the mass above the threshold. Because for a state above the threshold the relation $g^2=\frac{2\pi\sqrt{2\mu{B}}}{\mu^2}(1-Z)$ does not hold anymore, and the coupling between the compact state and the continuum states can be absorbed into the width term $\Gamma$ of the Breit-Wigner. That means, if one uses $G_X(E)$ to fit some near threshold structure, and the best fit gives $B<0$(above the threshold), then the meaningful fitting result should have $Z\rightarrow1$ at the same time. In such a case one can claim that the near threshold structure is a compact state with the mass above the threshold. Therefore $G_X(E)$ can also be used to identify the structure of threshold state with the mass above the threshold.

Secondly, for $Z=0$, by using the relation in Eq\eqref{sec:2}, $D_{EFT}(E)$ is equal to $-\dfrac{\sqrt{2\mu{B}}}{\mu}D_{LE}(E)$, with $D_{LE}(E)$ defined in the low-energy amplitude, i.e. Eq.\eqref{sec1:3}(In other words, low-energy amplitude can only be used for a pure molecule.).  This point has already been mentioned in Ref~\cite{Chen:2013upa}, we address it again for the completion of our discussions. Finally, for $Z\neq1$ and $Z\neq0$, $B_0$ and $g_0$ are well-defined. The form of the denominator of Eq.\eqref{sec3:5}, i.e., the propagator in its bare form, is precisely $D_{FL}(E)$ as defined in Eq.\eqref{sec1:2}. The matching between them yields the relations: $E_f=-B_0$, $g_1=g_0^2\mu/\pi$, $\Gamma_f=\Gamma_0$. By the way, the first two have been given in Ref~\cite{Chen:2013upa} and the last one is new. It can be observed that as $Z$ approaches 0, the Flatt$\acute{e}$ parameters ($E_f,g_1$) become infinite, due to ($B_0,g_0$) becoming infinite. Conversely, as $Z$ approaches 1, $g_1$ becomes 0, coinciding with the vanishing of $g_0^2$. Therefore, the Flatt$\acute{e}$ parameterization can only be successfully applied in the case $0<Z<1$. Hence the Flatt$\acute{e}$ amplitude assumes the existence of a compact object. 

Thus far, we have demonstrated that Breit-Wigner amplitude assumes a compact state lies below the threshold and weakly couples to the continuum state ($Z\rightarrow1$) or a compact state lies above threshold without knowing the coupling strength between this state and the continuum state. The low energy amplitude proposed in Ref\cite{Braaten:2007dw} assumes the below threshold state to be a pure molecule. The Flatt$\acute{e}$ amplitude assumes a below threshold compact state with its coupling strength with the continuum state determined by the value of $Z$. Therefore, all the three amplitudes have assumptions. In contrast, the propagator $G_X(E)$ derived from CEFT includes the factor $Z$ explicitly, thus it makes no assumptions on the structure of the near threshold states and provides a more general formula to describe S-wave threshold states.

We now come to further discuss the $X(3872)$, which may also consider the charged  $DD$ channel. With Fig.\ref{tu2}, the full propagator, which include the charged $DD$ channel, can be written as
\begin{equation}
G_{X(3872)}=\frac{i}{E+B_0-g_0^2\frac{\mu}{2\pi}\sqrt{-2\mu{E}-i\epsilon}-g_{0c}^2\frac{\mu_c}{2\pi}\sqrt{-2\mu_c(E-\delta)-i\epsilon}+i\Gamma_0/2}.  \label{Gx3872}
\end{equation}
Using the matching relations $g_0=\frac{1}{\sqrt{Z}}g$ (i.e. Eq.\eqref{sec2:07}), $g_{0c}=\frac{1}{\sqrt{Z}}g_c$ (i.e. Eq.\eqref{sec2:10}), and Eq.\eqref{sec2:26}, we can have 
\begin{equation}
G_{X(3872)}=\frac{iZ}{E+B+\widetilde{\Sigma^\prime}_{X(3872)}+i\Gamma/2},  \label{sec3:10}
\end{equation}
where
\begin{eqnarray}
\widetilde{\Sigma^\prime}_{X(3872)}&=&g^2\frac{\mu}{2\pi}(\sqrt{2\mu B}-\sqrt{-2\mu{E}-i\epsilon}) \nonumber\\
&&+g_c^2\frac{\mu_c}{2\pi}(\sqrt{2\mu_c(B+\delta)}-\sqrt{-2\mu_c(E-\delta)-i\epsilon})-(1-Z)(E+B)  \nonumber\\
&=&-g^2[\frac{\mu}{2\pi}\sqrt{-2\mu{E}-i\epsilon}+\frac{\mu\sqrt{2\mu B}}{4\pi B}(E-B)] \nonumber\\
&&-g_c^2[\frac{\mu_c}{2\pi}\sqrt{-2\mu_c(E-\delta)-i\epsilon}+\frac{\mu_c\sqrt{2\mu_c(B+\delta)}}{4\pi(B+\delta)}(E-B-2\delta)],  \label{newGx3872}
\end{eqnarray}
and $\Gamma=Z\Gamma_0$. Note that $g^2$ is determined by Eq.\eqref{sec2:18} instead of Eq.\eqref{sec2:1}, and we replace the factor $(1-Z)$ using Eq.\eqref{sec2:18} in the last step of Eq.\eqref{newGx3872}. One can also reobtain Eq.\eqref{sec2:26} by noting that $E=-B$ is the pole of Eq.\eqref{Gx3872} in the limit $\Gamma_0\rightarrow0$. It is worth mentioning that the Flatt$\acute{e}$ amplitude used by LHCb \cite{LHCb:2020xds} for the $X(3872)$ (which was first proposed in Ref\cite{Hanhart:2007yq}) is essentially the same as Eq.\eqref{Gx3872}. As previously emphasized, using the Flatt$\acute{e}$ amplitude involves an assumption, as the Flatt$\acute{e}$ amplitude parameters $B_0$, $g_0$, $g_{0c}$ become infinity in the limit $Z\rightarrow0$. However, with the renormalized form, i.e., Eq.\eqref{sec3:10}, to parameterize the propagator of the $X(3872)$, no assumption is made about the structure of $X(3872)$ since all quantities are finite for $0\leq Z\leq1$. Another important point is that by using the coupled channel Flatt$\acute{e}$ amplitude to extract $Z$, LHCb further uses the compositeness relations Eq.\eqref{sec:2} and Eq.\eqref{sec:3} (or equally, the relation between $Z$ and the asymmetry of the pole location in momentum space\cite{Baru:2003qq}). However these compositeness relations are only valid in the single -channel scattering. Therefore, such analysis relies on the extrapolation to the single-channel case, and it is unclear whether this extrapolation is valid. In contrast, one can directly determine the value of $Z$ by using Eq.\eqref{sec3:10} to fit the lineshape without any extrapolation, since Eq.\eqref{sec3:10} is derived from a coupled channel analysis.

Finally, we would like to discuss implications of our results in phenomenological studies. Considering a scattering process: initial states $\rightarrow$ $X$+others $\rightarrow$ final states, the scattering amplitude can be generally written as $i\mathcal{M}=\mathcal{A}_{PX}G_X\mathcal{B}_{DX}$, where $\mathcal{A}_{PX}$ is the production vertex of $X$, $G_X$ is the propagator of $X$ defined in Eq.\eqref{sec3:1}, and $\mathcal{B}_{DX}$ is the decay vertex of $X$. The production vertex of $X$ can be further separated into short-distant part $\mathcal{A}_{PX}^S$ and long-distant part $\mathcal{A}_{PX}^L$, i.e., ${A}_{PX}=\mathcal{A}_{PX}^S+\mathcal{A}_{PX}^L$. In the short-distant part $\mathcal{A}_{PX}^S$, $X$ is produced directly, while in the long-distant part $\mathcal{A}_{PX}^L$, $X$ is produced through the $DD$ rescattering. Specifically, if $X$ is a tetraquark (for $X(3872)$, it can also be $c\bar c$ ). $\mathcal{A}_{PX}^S$ can be estimated by the tetraquark model, i.e., no additional factors of $Z$ are needed to included in $\mathcal{A}_{PX}^S$. This can be immediately realized by replacing the propagator of $X$ in the Feynman diagram of the scattering process with the bubble sum in FIG.1 and notice that in the production vertex a bare state $|\mathcal{B}\rangle$ was first produced and then renormalized by the interaction. Similarly, if we consider the non-$DD$ decay modes of $X$, $\mathcal{B}_{DX}$ can be estimated by the tetraquark model. Note that in the way, we assume the non-$DD$ decay modes come from the direct decay of the tetraquark, instead of through the final state interaction of $DD$ rescattering. This assumption is justified for the radiative decay of $X(3872)$~\cite{Grinstein:2024rcu}. For the $X(3872)$, it is also found that $\mathcal{A}_{PX}^S>>\mathcal{A}_{PX}^L$~\cite{Duan:2024zuo,Bignamini:2009sk,Chen:2013upa}, so the scattering amplitude is dominated by $i\mathcal{M}=\mathcal{A}_{PX}^SG_X\mathcal{B}_{DX}$. Therefore, if we consider the non-$DD$ decay modes, $\mathcal{A}_{PX}^S$ and $\mathcal{B}_{DX}$ can be estimated by the tetraquark modes and the factor $Z$ only manifests itself in $G_X$ of the scatting amplitude. Otherwise, if we consider the $DD$ decay modes, the dominated scattering amplitude is written as $i\mathcal{M}=\mathcal{A}_{PX}^SG_Xig_0$. It is then interesting to constrain the parameters of the tetraquark models by fitting the cross section data with the above amplitudes. A non-vanishing $Z$ makes $\mathcal{A}_{PX}^S$ and non-$DD$ decay vertex $\mathcal{B}_{DX}$ can be estimated with the tetraquark model. This feature ensure that the large production cross section of $X(3872)$ at CDF~\cite{Bignamini:2009sk} and the large $R_{\psi\gamma}$ measured by LHCb~\cite{LHCb:2024tpv} and also strong coupling of $X(3872)$ with $D\bar D^*$ can be well described in a compact state model for $X(3872)$. This picture is consistent with the result of Ref\cite{Chen:2017tgv}, which found that while the nucleon-nucleon interaction can form a pure molecular state, the meson-meson interaction cannot form such a pure molecular state in large $N_c$ QCD. This discrepancy arises because the nucleon mass and the meson mass have different $N_c$ scaling.

\section{Summary}

We have re-derived the propagator for the near threshold states in CEFT. In this way, the $\Gamma$ can be naturally included in the propagator and can be interpreted as the renormalized non-$DD$ decay width of the compact state. We further discuss the relations between the propagator in CEFT and the Breit-Wigner, Flatt$\acute{e}$ and low-energy amplitude. We find that the propagator in CEFT is the more general formula to describe the S-wave near threshold states. By using this propagator to fit the lineshape, one can extract Z for these states and eventually clarify their structure. A study in this direction is presented in Ref~\cite{Xu:2023lll}. Finally, we discuss the implication of our results in the phenomenological studies.


\section{Acknowledgments}

Guoying Chen would like to thank Profs. Jianping Ma and Hanqing Zheng. Some of the ideas in the manuscript was inspired by the discussions with them during the past about twenty years. He would also like to thank Prof. Qiang Zhao for previous collaboration on exotic states. We also thank the anonymous referees for their valuable comments which help us to have a better understanding on Weinberg's compositeness theorem and motivate us to further consider the charged channel of $X(3872)$.
\bibliography{apssamp}

\end{document}